\documentclass[aps,preprint,floatfix,showpacs,letterpaper]{revtex4-1}
\usepackage[pdftex]{color}
\usepackage{latexsym,amssymb,amsmath,gensymb}
\usepackage{graphicx,dcolumn,bm}
\usepackage{slashed,kec,bigints}
\usepackage{mathtools}
\begin{document}

\title{New dates for the eras of the universe from the Planck data}
\author{Kevin Cahill}
\email{cahill@unm.edu}
\affiliation{Department of Physics \& Astronomy,
University of New Mexico, Albuquerque, New Mexico, USA}
\affiliation{School of Computational Sciences,\\
Korea Institute for Advanced Study, Seoul, Korea}
\date{\today}
\begin{abstract}
Data from the Planck satellite imply new dates for the major eras of the universe.   The era of radiation ended 50,150 years after inflation and the era of matter 10.31 billion years after inflation or 3.51 billion years ago.
\end{abstract}

\maketitle

\section{The Planck Data
\label{The Planck Data}}

The Planck Collaboration
and the European Space Agency have
published their remarkable data in a 
paper on 
the cosmological parameters~\cite{PlanckCosmological} 
and in 31 other arXiv articles.
These data imply that the era of radiation
ended 15 thousand years earlier 
and the era of matter ended
1.5 Gyr later than the WMAP data
had suggested.
\par
The Planck Collaboration used 
\( T_0 = 2.7255 \pm 0.0006 \) K~\cite{Fixsen2009}
as the current temperature of the (known) universe
and listed in column 7 of Table 2 
of their paper on 
cosmological parameters~\cite{PlanckCosmological}
these data:
The Hubble constant  is \(H_0 = 67.3 \pm 1.2\)
km/sMpc;   the fraction of the critical energy density 
\( \rho_c = 3 H_0^2/8\pi G = 8.506 \by 10^{-30} \) g cm\(^{-3}
= [ 0.002461 \mbox{ eV}]^4\)
that is due to dark energy is
\( \Omega_\Lambda = 0.685^{+0.018}_{-0.016} \);
the fraction due to matter is
\( \Omega_m = 0.315^{+0.016}_{-0.018} \);
and the  fraction due to ordinary matter
is \( \Omega_b = 0.0487 \pm .0006 \)
or 15.5\% of the total matter fraction.
The ratio of the energy density to the 
critical energy density is
\( \Omega = 1.000 \pm 0.036 \)\@.
 The age of the known universe
is \(13.817 \pm 0.048\) Gyr\@.
\par
I use these data and some standard formulas
to estimate first
in section~\ref{The End of the Era of Matter}
the time at which the energy
density of matter was the same as that
of dark energy and then
in section~\ref{The End of the Era of Radiation}
the time at which the energy
density of matter was the same as that
of radiation.
Both computations are straightforward
exercises.
The paper ends with a display
and brief discussion 
of two remarkable Planck 
CMB figures. 

\section{The End of the Era of Matter
\label{The End of the Era of Matter}}

Baryonic matter, mostly hydrogen
and helium, is stable on a time scales
that are much longer than the age of the universe. 
To estimate the end of the era
of matter, I will assume that dark matter also 
is stable on time scales of tens
of billions of years
(or that it decays into other forms
of matter)\@.
In this case,
the energy density \( \rho_m \) of all matter
and the scale factor \( a \) 
at any given time (after inflation)
are related
to their current values 
\( \rho_{m,0} \)  and \( a_0 \) by
\begin{equation}
a^3 \, \rho_m = a_0^3 \, \rho_{m,0} .
\label{energy density matter}
\end{equation}
I assume that the density of dark energy
is constant 
\begin{equation}
\rho_v = \rho_{v, 0} .
\label{density of dark energy}
\end{equation}
The era of matter ended when
\( \rho_m = \rho_v \),
which occurred when the
scale factor satsified
\begin{equation}
\rho_m = a_0^3 \, \rho_{m,0} /a^3 = \rho_v
\label{end of matter era}
\end{equation}
or
\begin{equation}
\frac{a}{a_0} = \lt( \frac{\rho_m}{\rho_v} \rt)^{1/3}
= \lt( \frac{\Omega_m}{\Omega_\Lambda} \rt)^{1/3}
= \lt( \frac{0.315}{0.685} \rt)^{1/3} = 0.7719 .
\label{a/a_0}
\end{equation}
This is a redshift of
\begin{equation}
z = a_0/a - 1 = 0.2956.
\label{z at end of matter era}
\end{equation}
\par
To convert this redshift
into a time, we use Eq.(1.5.42)
of Weinberg's \textsl{Cosmology}~\cite{Weinberg2010}
\begin{equation}
t(z) = \frac{1}{H_0} \int_0^{a/a_0}
\frac{1}{\sqrt{\Omega_\Lambda + \Omega_k x^{-2}
+ \Omega_m x^{-3} + \Omega_r x^{-4}}}
\frac{dx}{x} 
\label{1.5.42}
\end{equation}
in which \( \Omega_k = -K/(a_0 \, H_0)^2 \)
(with \( K = -1 \), 0, or 1)
is effectively zero.  
By Eq.(2.1.10) of his \textsl{Cosmology},
the fraction \( \Omega_r \)
of the energy density that is due to radiation is 
\begin{equation}
 \Omega_r = 4.15 \by 10^{-5} \, h^{-2}
  = 4.15 \by 10^{-5} / (0.673)^2 = 9.1626 \by 10^{-5} .
 \label{Omegar}
\end{equation}
Doing the integral (\ref{1.5.42}), we find
that the era of matter ended at about
\begin{equation}
t(0.2956) = 10.309 \mbox{ Gyr} 
\label{10.31}
\end{equation}
after inflation,
give or take a few million years.
Thus the era of dark energy began 
about 3.508 billion years ago.

\section{The End of the Era of Radiation
\label{The End of the Era of Radiation}}

I will take the energy of radiation as
due to photons and nearly massless neutrinos
whose wavelengths are stretched by
the scale factor \(a\)\@.
Their energy density
\( \rho_r \) then is related to its current value
\( \rho_{r,0} \) by
\begin{equation}
a^4 \,  \rho_r = a_0^4 \,  \rho_{r,0} .
\label{energy density radiation}
\end{equation}
The era of radiation ended when
\begin{equation}
\rho_r = a_0^4 \, \rho_{r,0}/a^4 = \rho_m = 
a_0^3 \, \rho_{m,0} /a^3 
\label{era of radiation ended}
\end{equation}
or when
\begin{equation}
\frac{a}{a_0} = \frac{\rho_{r,0}}{\rho_{m,0}}
= \frac{\Omega_r}{\Omega_m}
= \frac{9.1626 \by 10^{-5}}{0.315} = 2.9088 \by 10^{-4}
\label{a/a0 end of radiation}
\end{equation}
which is a redshift of
\begin{equation}
z = a_0/a -1 = 3436.9 .
\label{redshift end of radiation}
\end{equation}
Using the value (\ref{a/a0 end of radiation})
as the upper limit
of the integral (\ref{1.5.42}), 
we get as
the end of the era of radiation 
\begin{equation}
t(3436.9) = 50,152 \mbox{ yr}
\end{equation}
after inflation,
give or take a few decades.

\section{Two Images
\label{Two Images}}

\begin{figure}
\centering
\includegraphics[width=5.9in]{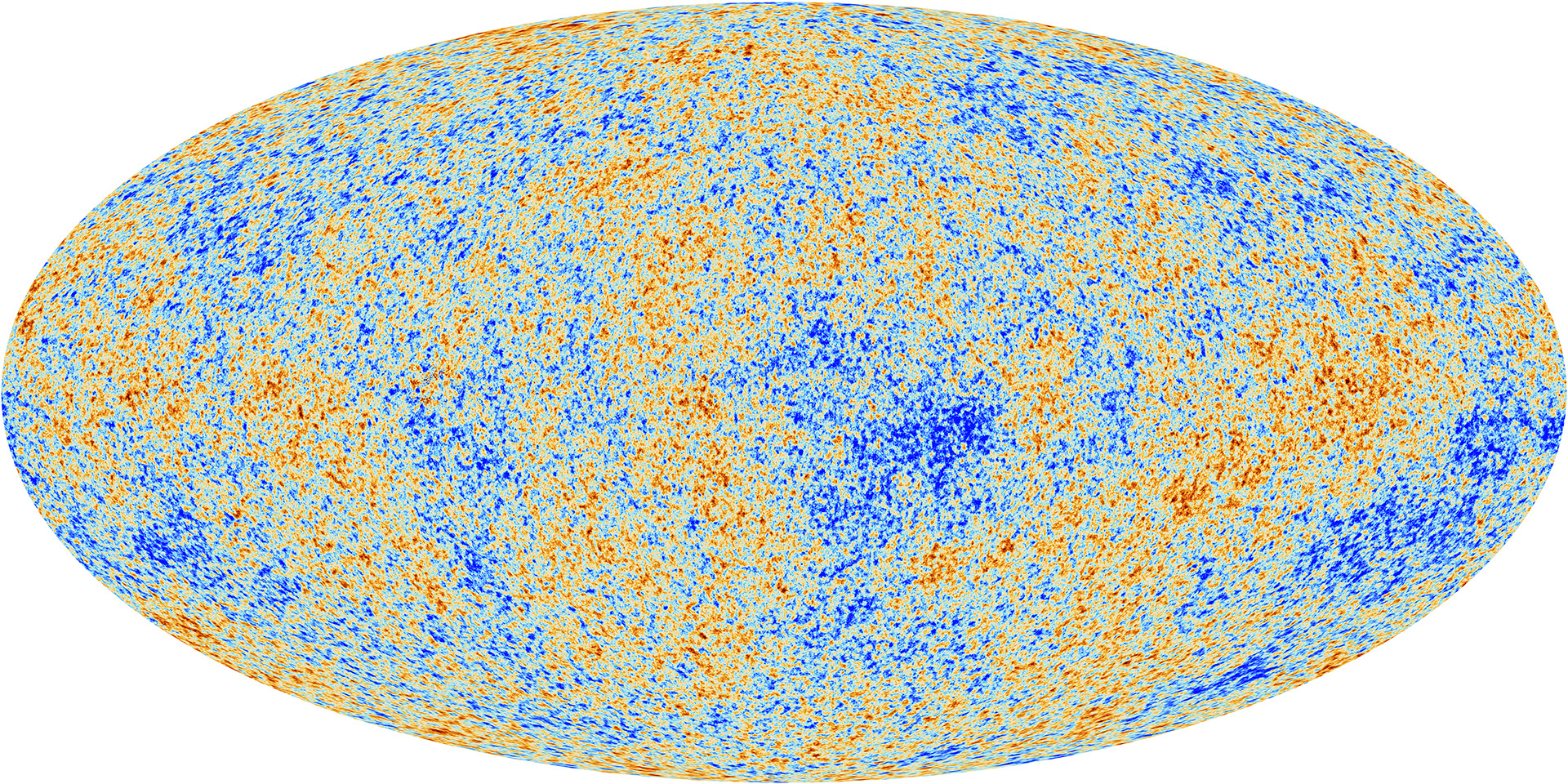}
\caption{CMB temperature fluctuations 
over the celestial sphere
as measured by the Planck satellite.  
The average temperature is 2.7255 K\@. 
Red regions are warmer and 
blue ones colder by \( \lesssim 500 \, \mu\)K\@. 
\copyright \, ESA and the Planck Collaboration.}
\label {CMB face of god}
\end{figure}

I cannot resist displaying
two extraordinary figures
published by the Planck Collaboration.
The first is their full-sky image of the 
tiny fluctuations in the
temperature of the
cosmic microwave background radiation.
In this Fig.~\ref{CMB face of god},
red is hotter and blue colder
by \(  \lesssim 500 \, \mu\)K\@. 

\begin{figure}
\centering
\includegraphics[width=5.9in]
{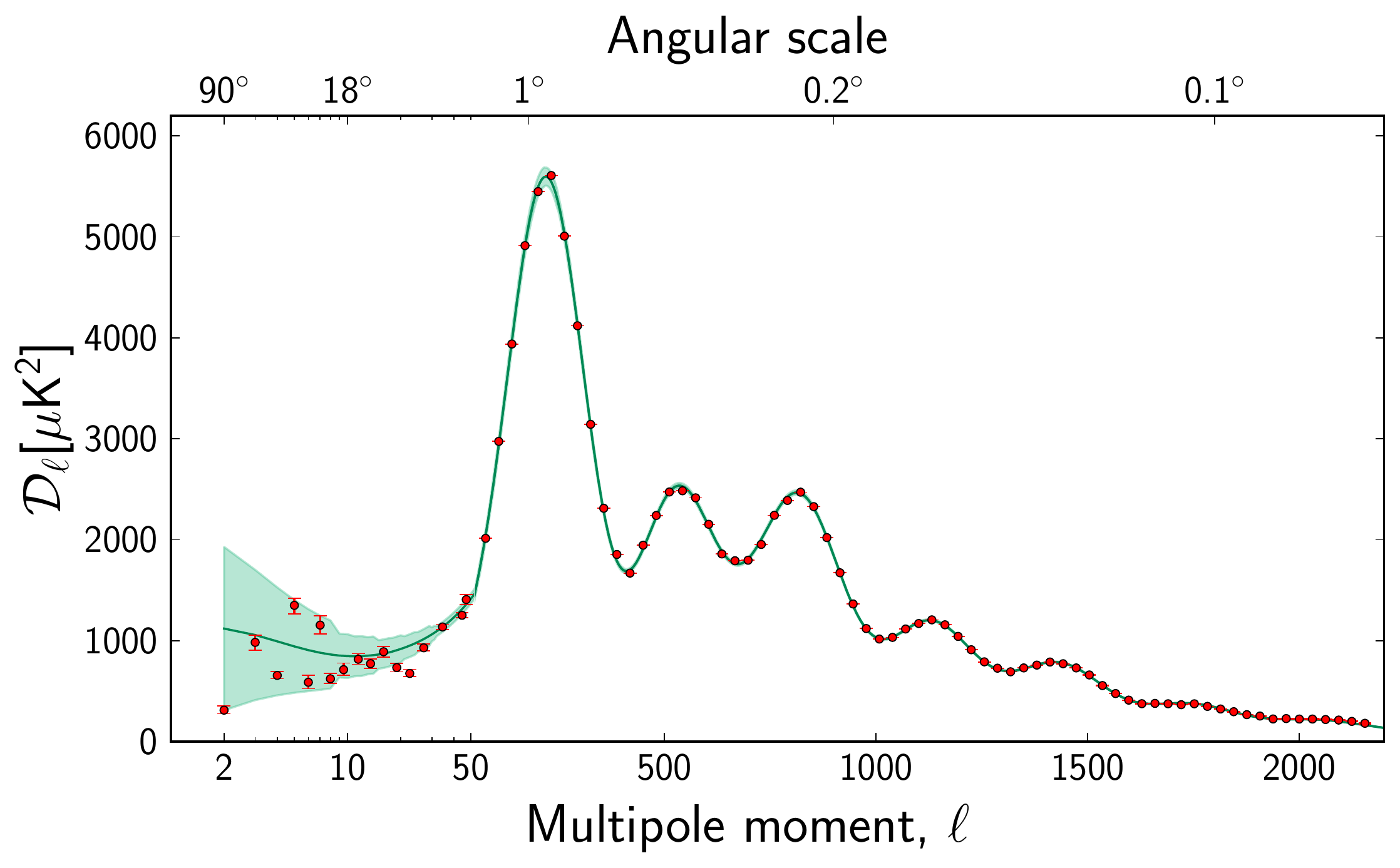}
\caption{The power spectrum 
\(\mathcal{D}_\ell = \ell(\ell+1) C_\ell/2\pi\)
of the CMB temperature fluctuations
in \( \mu \mbox{K}^2 \)
as measured by the Planck
Collaboration (arXiv:1303.5062)
is plotted against the angular size 
and the multipole moment \(\ell\)\@.
The solid curve is the \(\Lambda\)CDM prediction.}
\label {CMB l plot}
\end{figure}

The second remarkable figure 
is their graph of the 
angular correlations of 
the full-sky, CMB temperature data
shown in Fig.~\ref{CMB l plot}\@.
They expanded the temperature \( T(\theta,\phi) \) 
in spherical harmonics as~\cite{CahillVIII}
\beq
T(\th,\phi) = \sum_{\ell = 0}^\infty
\sum_{m=-\ell}^\ell a_{\ell,m} \, Y_{\ell, m}(\th,\phi) 
\label {T expansion}
\eeq
in which the coefficients are 
\beq
a_{\ell,m} = \int_0^{2\pi} \!\!d\phi \int_0^\pi \!\! 
\sin\theta \, d\theta \,\,\,
Y^*_{\ell, m}(\theta,\phi) \, T(\theta,\phi).
\label {alm =}
\eeq
The average temperature \(\overline T\) contributes
only to \(a_{0,0} = \overline T = 2.7255 \) K\@.
The other coefficients describe 
the difference
\(\Delta T(\th,\phi) = T(\th,\phi) - \overline T\)\@.
The angular power spectrum is 
\beq
C_\ell = \frac{1}{2\ell + 1} 
\sum_{m=-\ell}^\ell |a_{\ell,m}|^2.
\label {T power spectrum}
\eeq
They displayed these coefficients scaled
by \(\ell(\ell+1)/2\pi \) in Fig.~\ref{CMB l plot}\@.
\par
This power spectrum is a snapshot
at the moment of initial transparency
(380,000 years after inflation)
of the temperature distribution
of the plasma of photons, electrons, and nuclei
undergoing acoustic oscillations.
In these oscillations,
gravity opposes radiation pressure,
and  \( |\Delta T(\theta,\phi) | \) is maximal
both when the oscillations are most compressed and 
when they are most rarefied.
Regions that gravity has been squeezing
over the first 330,000 years
of the era of matter and
that reach maximum compression at
the moment of initial transparency
form the first and highest peak.
Regions that have bounced off their 
first compression and that
radiation pressure has expanded
to minimum density at initial transparency
form the second peak.
Regions that reach maximum
compression for a second time at transparency
form the third peak, and so forth.
The agreement of the \(\Lambda\)CDM
model with the Planck data 
is excellent for \( \ell > 40 \)\@.
Discrepancies below \( \ell = 40 \)
may point to new physics.

\begin{acknowledgments}
I should like to thank
Jooyoung Lee for inviting
me to KIAS and Rouzbeh
Allahverdi and Franco Giuliani
for several helpful conversations.
\end{acknowledgments}
\bibliography{physics,astro}
\end{document}